\documentclass{article}%
\usepackage{amsfonts}
\usepackage{amsmath}
\usepackage{amssymb}
\usepackage{graphicx}%
\begin{document}

\title{Equivalence Principle and the Principle of Local Lorentz
Invariance\thanks{published: \emph{Found. of Physics} \textbf{31},
1785-1806 (2001). Includes corrigenda published \emph{Found.
Phys.} \textbf{32}, 811-812 (2002).}}
\author{\quad\ W. A. Rodrigues, Jr.$^{(1,2)}\hspace{0.01in}$\thanks{e-mail:
walrod@ime.unicamp.br } and M. Sharif$^{(1)}$\thanks{Permanent Address:
Department of Mathematics, Punjab University, Quaid-e-Azam Campus
Lahore-54590, PAKISTAN, e-mail: hasharif@yahoo.com}\\$^{(1)}${\footnotesize Institute of Mathematics, Statistics and Scientific
Computation }\\{\footnotesize \ IMECC--UNICAMP CP 6065}\\{\footnotesize \ 13083-970 Campinas, SP, Brazil}\\{\footnotesize \ and}\\{\footnotesize \ }$^{(2)}${\footnotesize Wernher von Braun}\\{\footnotesize \ Advanced Research Center, UNISAL}\\{\footnotesize \ Av. A. Garret, 257}\\{\footnotesize \ 13087-290 Campinas, SP Brazil}\\pacs: 04.90+e \ 03.30+p}
\date{02/04/2003}
\maketitle
\tableofcontents

\begin{abstract}
In this paper we scrutinize the so called Principle of Local Lorentz
Invariance (\emph{PLLI}) that many authors claim to follow from the
Equivalence Principle. Using rigourous mathematics we introduce in the General
Theory of Relativity two classes of reference frames (\emph{PIRFs} and
\emph{LLRF}$\gamma$\emph{s}) which natural generalizations of the concept of
the inertial reference frames of the Special Relativity Theroy. We show that
it is the class of the \emph{LLRF}$\gamma$\emph{s} that is associated with the
\emph{PLLI.} Next we give a defintion of physically equivalent referefrence
frames. Then, we prove that there are models of General Relativity Theory (in
particular on a Friedmann universe) where the \emph{PLLI }is false. However
our find is not in contradiction with the many experimental claims vindicating
the \emph{PLLI}, because theses experiments do not have enough accuracy to
detect the effect we found. We prove moreover that \emph{PIRFs }are not
physically equivalent.

\end{abstract}

\section{Introduction}

In this paper we scrutinize the so called Principle of Local Lorentz
Invariance (\emph{PLLI}) that some authors claim to follow from the
Equivalence Principle (\emph{EP}). We show that \emph{PLLI} is false according
to General Theory of Relativity (\emph{GRT}), but nevertheless $it$ is a very
good approximation in the physical world we live in. In order to prove our
claim, we recall the mathematical definition of reference frames in \emph{GRT
} which are modelled as certain unit timelike vector fields. We study the
classification of reference frames and give a physically motivated and
mathematical rigorous definition of physically equivalent reference frames. We
investigate next which are the reference frames in \emph{GRT} which share some
of the properties of the inertial reference frames (\emph{IRFs}) of the
Special Theory of Relativity (\emph{SRT)}. We found that there are two kind of
frames that appear as generalizations of the \emph{IRFs} of \emph{SRT}. These
are the pseudo inertial reference frames (\emph{PIRFs}) \ and the local
Lorentz reference frames (\emph{LLRF}$\gamma$\emph{s)}. Now, \emph{PLLI} is a
statement that \emph{LLRF}$\gamma$\emph{s} are physically equivalent. We show
that \emph{PLLI} is false by expliciting showing that there are models of
\emph{GRT} (explicitly a Friedmann Universe) containing \emph{LLRF}$\gamma
$\emph{s} which are not physically equivalent.

We emphasize that our finding is not in contradiction with the many
experimental proofs offered as vindicating the \emph{PLLI}, since all this
proofs do not have enough accuracy to detect the effect we have found which is
proportional to $2av^{2}$, where $a<<1$ and $v$ is the initial metric velocity
of the \emph{LLRF}$\gamma^{\prime}\mathbf{L}^{\prime}$ in relation to a
\emph{LLRF}$\gamma$ $\mathbf{L}$ in a Friedmann Universe model of \emph{GRT}.
Indeed, for this model we showed that any point of the world manifold $p$
there is a \emph{LLRF}$\gamma$ $\mathbf{L}$\textbf{, }for which $\mathbf{V|}%
_{\gamma}=\mathbf{L|}_{\gamma}$ ( where $\mathbf{V}$\textbf{\ }is a
fundamental \emph{PIRF} such that the center of mass of each galactic cluster
follows one of its integral lines) and such that its expansion ratio at $p$ is
null and that if $\mathbf{L}^{\prime}$\emph{\ }is a \emph{LLRF}$\gamma
^{\prime}$ ($\gamma\cap\gamma^{\prime}=p\in M$) moving with initial metric
velocity $v$ at $p$ relative $\mathbf{L}$ then the expansion ratio of
$\mathbf{L}^{\prime}$ is $2av^{2}$.

We prove also that there are models of \emph{GRT} where \emph{PIRFs }are not
physically equivalent also.

\section{Some Basic Definitions}

\subsection{Relativistic Spacetime Theories}

In this subsection we recall what we mean by a relativistic spacetime theory
[1], a key concept necessary to prove our claim that the so called \emph{PLLI}
is not a fidedigne law of nature.

In our approach a physical theory $\tau$ is characterized by:

{(i)} a theory of a certain \emph{species of structure} in the sense of
Boubarki [2];

{(ii)} its physical interpretation;

{(iii)} its present meaning and present applications.

We recall that in the mathematical exposition of a given physical theory
$\tau$, the postulates or basic axioms are presented as definitions. Such
definitions mean that the physical phenomena described by $\tau$ behave in a
certain way. Then, the definitions require more motivation than the pure
mathematical definitions. We call coordinative definitions the physical
definitions, a term introduced by Reichenbach [3]. Also, according to Sachs
and Wu [4] it is necessary to make clear that completely \emph{convincing} and
\emph{genuine} motivations for the coordinative definitions cannot be given,
since they refer to nature as a whole and to the physical theory as a whole.

The theoretical approach to physics behind (i), (ii) and (iii) above is then
to admit the mathematical concepts of the \emph{species of structure} defining
$\tau$ as primitives, and define coordinatively the observation entities from
them. Reichenbach assumes that ``\emph{physical knowledge} is characterized by
the fact that concepts are not only defined by other concepts, but are also
coordinated to real objects''. However, in our approach, each physical theory,
when characterized as a species of structure, contains some \emph{implicit
}geometric objects, like some of the reference frame fields defined below,
that cannot in general be coordinated to real objects. \footnote{Indeed, it
would be an absurd to suppose that all the infinity of \emph{IRFs
}(observation 3) that exist in a Minkowski spacetime are simultaneously
realized as physical systems.}

\textbf{Definition 1}. A general relativistic \emph{spacetime} theory is as a
theory of a species of structure such that, if Mod $\tau$ is the class of
models of $\tau$, then each $\Upsilon\in$ Mod $\tau$ contains as substructure
a Lorentzian spacetime $ST=\langle M,D,g\rangle$. We recall here that $g$ is a
Lorentz metric and $D$ is the Levi-Civita connection of $g $ on $M$ [4]. More
precisely, we have
\begin{equation}
\Upsilon=(M,D,g,\mathbf{T}_{1},\ldots,\mathbf{T}_{m})\;, \label{4.12BIS}%
\end{equation}
The $\mathbf{T}_{i}\in\sec\tau M$ (the tensor bundle), $i=1,\ldots,m$ are
(\emph{explicit}) geometrical objects defined in $U\subseteq M$ characterizing
the physical fields and particle trajectories that \emph{cannot} be
geometrized in the theory. Here, to be geometrizable means to be a metric
field or a connection on $M$ or objects derived from these concepts as,
\emph{e.g.\/}, the Riemann tensor (or the torsion tensor in more general
spacetime theories). The $\mathbf{T}_{i}$, $i=1,\ldots,m$ are supposed to
satisfy a set of differential equations involving also $D$ and $g$ called the
dynamical laws of the theory.

As already said above, each spacetime theory has some \textit{implicit}
geometrical that do not appear explicitly in eq.(\ref{4.12BIS}). These objects
are the reference frame fields which we now study and analyze in detail.

\subsection{Reference Frames}

\textbf{Definition 2.} Let $ST$ be a relativistic spacetime. A \emph{moving
frame} at $x\in M$ is a basis for the tangent space $T_{x}M$. An
\emph{orthonormal frame} for $x\in M$ is a basis of orthonormal vectors for
$T_{x}M$.

\textbf{Proposition 3.} Let $\mathbf{Q}\in\sec TU\subset\sec TM$ be a
time-like vector field such that $\mathbf{g}(\mathbf{Q},\mathbf{Q})=1$. Then,
there exist, in a coordinate neighborhood $U$, three space-like vector fields
which together with $\mathbf{Q}$ form an orthogonal moving frame for $x\in U$.
The proof is trivial [7].

\textbf{Definition 2.} A non-spinning particle on $ST$ is a pair $(m,\sigma)$
where $\sigma:\mathcal{R}\supset I\rightarrow M$ is a future pointing causal
curve [4-6] and $m\in[0,+\infty)$ is the mass. When $m=0$ the particle is
called a photon. When $m\in(0,+\infty)$ the particle is said to be a material
particle. $\sigma$ is said to be the world line of the particle.

\textbf{Definition 3.} An \emph{observer} in $<M,D,\mathbf{g}>$ is a future
pointing time-like curve $\gamma:\mathcal{R}\supset I\rightarrow M$ such that
$\mathbf{g}(\gamma_{*}u,\gamma_{*}u)=1$. The inclusion parameter
$I\rightarrow\mathcal{R}$ in this case is called the proper time along
$\gamma$, which is said to be the world line of the observer.

\textbf{Observation 1.} The physical meaning of proper time is discussed in
details, e.g., in [5,6] which deals with the theory of time in relativistic theories.

\textbf{Definition 4.} An \emph{instantaneous observer} is an element of $TM$,
i.e., a pair $(z,\mathcal{Z}),$ where $z\in M$, and $\mathcal{Z}\in T_{z}M$ is
a future pointing unit time-like vector.

The Proposition 1 together with the above definitions suggests:

\textbf{Definition 5.} A \emph{reference frame} in $ST=<M,D,\mathbf{g}>$ is a
time-like vector field which is a section of $TU,U\subseteq M$ such that each
one of its integral lines is an observer.

\textbf{Observation 2.} In [4-6] an arbitrary reference frame $\mathbf{Q\in
\sec}TU\subseteq\sec TM$ is classified as (i), (ii) below.

(i) according to its \emph{synchronizability}$.$ Let $\alpha_{\mathbf{Q}%
}=g(\mathbf{Q},)$. We say that \textbf{Q} is locally synchronizable iff
$\alpha_{\mathbf{Q}}\wedge d\alpha_{\mathbf{Q}}=0$. \textbf{Q }is said to be
locally proper time synchronizable iff $d\alpha_{\mathbf{Q}}=0$. \textbf{Q }is
said to be synchronizable iff there are $C^{\infty}$ functions
$h,t:M\rightarrow\mathcal{R}$ such that $\alpha_{\mathbf{Q}}=hdt$ and $h>0$.
\textbf{Q} is proper time synchronizable iff $\alpha_{\mathbf{Q}}=dt$. These
definitions are very intuitive.

(ii) according to the \emph{decomposition} of
\begin{equation}
D\alpha_{\mathbf{Q}}=\mathbf{a}_{\mathbf{Q}}\otimes\alpha_{\mathbf{Q}%
}+\mathbf{\omega}_{\mathbf{Q}}+\mathbf{\sigma}_{\mathbf{Q}}+\frac{1}%
{3}\mathbf{\Theta}_{\mathbf{Q}}\mathbf{h}, \label{0}%
\end{equation}
where
\begin{equation}
\mathbf{h}=\mathbf{g}-\alpha_{\mathbf{Q}}\otimes\alpha_{\mathbf{Q}} \label{1}%
\end{equation}
is called the projection tensor (and gives the metric of the rest space of an
instantaneous observer [17-19]), $\mathbf{a}_{\mathbf{Q}}$ is the (form)
acceleration of \textbf{Q}, $\mathbf{\omega}_{\mathbf{Q}}$ is the rotation of
$\mathbf{Q}$, $\mathbf{\sigma}_{\mathbf{Q}}$ is the shear of $\mathbf{Q}$ and
$\mathbf{\Theta}_{\mathbf{Q}}$ is the expansion ratio of $\mathbf{Q}$ . In a
coordinate chart ($U,x^{\mu}$), writing $\mathbf{Q}=Q^{\mu}\partial/\partial
x^{\mu}$ \ and $\mathbf{h}=(g_{\mu\nu}-Q_{\mu}Q_{\nu})dx^{\mu}\otimes dx^{\nu
}$ we have
\begin{align}
\mathbf{\omega}_{\mathbf{Q}\mu\nu}  &  =Q_{\left[  \mu;\nu\right]
},\nonumber\\
\mathbf{\sigma}_{\mathbf{Q}\alpha\beta}  &  =[Q_{\left(  \mu;\nu\right)
}-\frac{1}{3}\mathbf{\Theta}_{\mathbf{Q}}h_{\mu\nu}]h_{\alpha}^{\mu}h_{\beta
}^{\nu},\nonumber\\
\mathbf{\Theta}_{\mathbf{Q}}  &  =Q^{\mu};_{\mu}. \label{2}%
\end{align}

We shall need in what follows the following result that can be easily proved:
\begin{equation}
\alpha_{\mathbf{Q}}\wedge d\alpha_{\mathbf{Q}}=0\Leftrightarrow\mathbf{\omega
}_{\mathbf{Q}}=0. \label{3}%
\end{equation}

Eq.(3) means that rotating reference frames (i.e., frames for which
$\mathbf{\omega}_{\mathbf{Q}}\neq0$) are not locally synchronizable. This
result is the key in order to solve the misconceptions usually associated with
rotating reference frames even in the \emph{SRT} (see [8] for examples).

\textbf{Observation 3}. In Special Relativity where the space time manifold is
$<M\mathcal{=R}^{4},\mathbf{g}=\mathbf{\eta},D^{\mathbf{\eta}}>$%
\footnote{$\mathbf{\eta}$ is a constant metric, i.e., there exists a chart
$\langle x^{\mu}\rangle$ of $\ M=\mathcal{R}^{4}$ such that $\mathbf{\eta
}(\partial/\partial x^{\mu},\partial/\partial x^{\nu})=\eta_{\mu\nu}$, the
numbers $\eta_{\mu\nu}$ forming a diagonal matrix with entries $(1,-1,-1,-1)$.
Also, $D^{\mathbf{\eta}}$ is the Levi-Civita connection of $\mathbf{\eta}$.}
an \emph{inertial reference frame }(\emph{IRF}) $\mathbf{I}\in\sec TM$ is
defined by $D^{\mathbf{\eta}}\mathbf{I}=0$. \ We can show very easily that in
\emph{GRT} \ where each gravitational field is modelled by a spacetime
$<M,\mathbf{g},D>$ there are no frame $\mathfrak{Q}\in$ $\sec TM$ satisfying
\ $D\mathfrak{Q}=0$. So, no \emph{IRF} exist in any model of \emph{GRT}%
.\medskip

The following question arises naturally: which characteristics a reference
frame on a \emph{GRT} spacetime model must have in order to reflect as much as
possible the properties of an \emph{IRF} of \emph{SRT}?

The answer to the question is that there are two kind of frames in \emph{GRT}
[\emph{PIRFs }(definition 6) and \emph{LLRFs }(definition 9)], such that each
frame in one of these classes share some important aspects of the \emph{IRFs}
of \emph{SRT}. Both concepts are important and as we will see, it is important
to distinguish between them in order to avoid misunderstandings.

\subsubsection{Pseudo Inertial Reference Frames}

\textbf{Definition 6.} A reference frame $\mathfrak{I}\in\sec TU,U\subset M$
is said to be a \emph{pseudo inertial} \emph{reference frame }(\emph{PIRF}) if
$D_{\mathfrak{I}}\mathfrak{I}=0$ and $\alpha_{\mathfrak{I}}\wedge
d\alpha_{\mathfrak{I}}=0, $ and $\alpha_{\mathfrak{I}}=g(\mathfrak{I},)$.

This definition means that a \emph{PIRF} is in free fall and is non rotating.
It means also that it is at least locally synchronizable.

\subsubsection{Naturally Adapted Charts to a Given Reference Frame}

\textbf{Definition 7.} Let $\mathbf{Q}\in\sec TU,U\subseteq M$ be a reference
frame. A chart in $U$ of the maximal oriented atlas of $M$ with coordinate
functions $\langle x^{\mu}\rangle$ such that $\partial/\partial x^{0}\in\sec
TU$ is a timelike vector field and the $\partial/\partial x^{i}\in\sec TU$
($i=1,2,3$) are spacelike vector fields is said to be a possible naturally
adapted coordinate chart to the frame \textbf{Q }(denoted $\emph{(nacs|}%
\mathbf{Q}\emph{)}$ in what follows) if the space-like components of
$\mathbf{Q}$ are null in the natural coordinate basis $\langle\partial
/\partial x^{\mu}\rangle$ of $TU$ associated with the chart.\footnote{We can
be prove very easily that there is an infinity of different $\emph{(nacs|}%
\mathbf{Q}\emph{).}$}

\subsubsection{Local Lorentzian Coordinate Chart}

\textbf{Definition 8.} A chart $(U,\xi^{\mu})$ of the maximal oriented atlas
of $M$ is said to be a \emph{local Lorentzian coordinate chart (LLCC) }and
$\langle\xi^{\mu}\rangle$ are said to be \emph{local Lorentz coordinates}
(\emph{LLC}) in $p_{0}\in U$ iff
\begin{equation}
\mathbf{g}(\partial/\partial\xi^{\mu},\partial/\partial\xi^{\nu})\mid_{p_{0}%
}=\eta_{\mu\nu}, \label{4}%
\end{equation}
\begin{equation}
\Gamma_{\beta\mu}^{\alpha}(\xi^{\mu})\mid_{p_{0}}=0,\quad\Gamma_{\beta
\gamma,\mu}^{\alpha}(\xi^{\mu})\mid_{p}=-\frac{1}{3}(R_{\beta\gamma\mu
}^{\alpha}(\xi^{\mu})+R_{\gamma\beta\mu}^{\alpha}(\xi^{\mu}))\mid_{p},\text{
}p\neq p_{0} \label{5}%
\end{equation}

Let $(V,x^{\mu})$ ($V\cap U\neq\emptyset$) be an arbitrary chart. Then,
supposing that $p_{0}$ is at the origin of both coordinate systems the
following relations holds (approximately)
\begin{align}
\xi^{\mu}  &  =x^{\mu}+\frac{1}{2}\Gamma_{\alpha\beta}^{\mu}(p_{0})x^{\alpha
}x^{\beta},\nonumber\\
x^{\mu}  &  =\xi^{\mu}-\frac{1}{2}\Gamma_{\alpha\beta}^{\mu}(p_{0})\xi
^{\alpha}\xi^{\beta}, \label{6}%
\end{align}
where in eqs.(\ref{6}) $\Gamma_{\alpha\beta}^{\mu}(p_{0})$ are the values of
the connection coefficients at $p_{0}$ expressed in the chart $(V,x^{\mu})$.

The coordinates $\langle\xi^{\mu}\rangle$ are also known as Riemann normal
coordinates and the explicit methods for obtaining them are described in many
texts of Riemaniann geometry as e.g., [9,10] and of \emph{GRT}, as e.g., [11,12].

\textbf{Observation 4.} Let $\gamma$ $\in U\subset M$ be the world line of an
observer in geodetic motion in spacetime, i.e., $D_{\gamma_{*}}\gamma_{*}=0 $.
Then as it is well known [11] we can introduce in $U$ a \emph{LLC} $\langle
\xi^{\mu}\rangle$ such that for every $p\in\gamma$ we have%

\begin{align}
\left.  \frac{\partial}{\partial\xi^{0}}\right|  _{p\in\gamma}  &
=\gamma_{*|p};\quad\left.  \mathbf{g}(\partial/\partial\xi^{\mu}%
,\partial/\partial\xi^{\nu})\right|  _{p\in\gamma}=\eta_{\mu\nu},\nonumber\\
\left.  \Gamma_{\nu\rho}^{\mu}(\xi^{\mu})\right|  _{p\in\gamma}  &  =\left.
g^{\mu\alpha}\mathbf{g}(\partial/\partial\xi^{\alpha},D_{\partial/\partial
\xi^{\nu}}\partial/\partial\xi^{\rho})\right|  _{p\in\gamma}=0. \label{7}%
\end{align}

Take into account for future reference that if the $<\xi^{\mu}>$ are \emph{LLC
}then it is clear from definition 8 that in general $\Gamma_{\mu\rho}^{\nu
}(\xi^{\mu})\mid_{p}\neq0$ for all $p\notin\gamma$.

\subsubsection{Local Lorentz Reference Frame}

\textbf{Definition 9}. Given a geodetic line $\gamma\subset U$ $\subset M$ and
\emph{LLCC }$(U,\xi^{\mu})$\emph{\ }we say that reference frame $\mathbf{L=}%
\partial/\partial\xi^{0}\in\sec TU$ is a \emph{Local Lorentz reference frame
associated to }$\gamma$ (\emph{LLRF}$\gamma$)\footnote{When no confusion
arises and $\gamma$ is clear from the context we simply write \emph{LLRF}.}
iff
\begin{align}
\left.  \mathbf{L}\right|  _{p\in\gamma}  &  =\left.  \frac{\partial}%
{\partial\xi^{0}}\right|  _{p\in\gamma}=\left.  \gamma_{*}\right|
_{p},\nonumber\\
\left.  \alpha_{\mathbf{L}}\wedge d\alpha_{\mathbf{L}}\right|  _{p\in\gamma}
&  =0. \label{8}%
\end{align}

Moreover, we say also that the Riemann normal coordinate functions or Lorentz
coordinate functions (\emph{LLC}) $<\xi^{\mu}>$ are \emph{associated} with the
\emph{LLRF}$\gamma.$

\textbf{Observation 5}. It is very important to have in mind that for a
\emph{LLRF}$\gamma$ \textbf{L} in general $\left.  D_{\mathbf{L}}%
\mathbf{L}\right|  _{p\notin\gamma}\neq0$ (i.e., only the integral line
$\gamma$ of \textbf{L} in free fall in general),\textbf{\ }and also eventually
$\left.  \alpha_{\mathbf{L}}\wedge d\alpha_{\mathbf{L}}\right|  _{p\notin
\gamma}\neq0$, which may be a surprising result for many readers. In contrast,
a \emph{PIRF} $\mathfrak{I}$ such that $\left.  \mathfrak{I}\right|  _{\gamma
}=\left.  \mathbf{L}\right|  _{\gamma}$ has all its integral lines in free
fall and the rotation of the frame is always null in all points where the
frame is defined. Finally its is worth to recall that both $\mathfrak{I}$ and
\textbf{L} may eventually have shear and expansion even at the points of the
geodesic line $\gamma$ that they have in common. This last point will be
important in our analysis of the \emph{PLLI} in section 6.

\textbf{Definition 10}. Let $\gamma$ be a geodetic line as in definition 9. A
section $s$ of the orthogonal frame bundle $FU,U\subset M$ is called an
\emph{inertial moving frame along }$\gamma$ $\emph{(IMF}\gamma\emph{)}$\ when
the set
\begin{equation}
s_{\gamma}=\{(e_{0}(p),e_{1}(p),e_{2}(p),e_{3}(p)),p\in\gamma\cap U\}\subset
s, \label{9}%
\end{equation}
it such that $\forall p\in\gamma$
\begin{equation}
e_{0}(p)=\left.  \gamma_{*}\right|  _{p},\left.  \mathbf{g}(e_{\mu},e_{\nu
})\right|  _{p\in\gamma}=\eta_{\mu\nu} \label{10}%
\end{equation}
with
\begin{equation}
\Gamma_{\nu\rho}^{\mu}(p)=g^{\mu\alpha}\mathbf{g}(e_{\alpha}(p),D_{e_{\nu}%
(p)}e_{\rho}(p))=0 \label{11}%
\end{equation}

\textbf{Observation 6.} The existence of $s\in\sec FU$ satisfying the above
conditions can be easily proved [9]. Introduce coordinate functions $<\xi
^{\mu}>$ for\emph{\ }$U$ such that at $p_{0}\in\gamma,e_{0}(p_{0})=\left.
\frac{\partial}{\partial\xi^{0}}\right|  _{p_{o}}=\gamma_{*|p_{0}}$, and
$e_{i}(p_{0})=\left.  \frac{\partial}{\partial\xi^{i}}\right|  _{p_{o}%
},i=1,2,3$ (three orthonormal vectors) satisfying Eq.(\ref{7} ) and parallel
transport the set $e{_{\mu}(}p{_{0})}$ along $\gamma$. The set ${e_{\mu}%
(}p{_{0})}$ will then also be \emph{Fermi} transported [4] since $\gamma$ is a
geodesic and as such they define the standard of \emph{no} rotation along
$\gamma.$

\textbf{Observation 7.} Let $\mathfrak{I}\in\sec TV$ be a \emph{PIRF} and
$\gamma$ $\subset U\subset V$ one of its integral lines and let $<\xi^{\mu}>,$
$U\subset M$ be a \emph{LLC} through all the points of the world line $\gamma$
such that $\gamma_{*}=\left.  \mathfrak{I}\right|  _{\gamma}$. Then, in
general $<\xi^{\mu}>$ is not a $(nacs|\mathfrak{I})$ in $U$, i.e., $\left.
\mathfrak{I}\right|  _{p\notin\gamma}\neq\left.  \partial/\partial\xi
^{0}\right|  _{p\notin\gamma}$ even if $\left.  \mathfrak{I}\right|
_{p\in\gamma}=\left.  \partial/\partial\xi^{0}\right|  _{p\in\gamma}$.

\textbf{Observation 8}. Before concluding this section it is very much
important to recall again that a reference frame field as introduced above is
a \emph{mathematical} instrument. It did not necessarily need to have a
material substratum (i.e., to be realized as a material physical system) in
the points of the spacetime manifold where it is defined. More properly, we
state that the integral lines of the vector field representing a given
reference frame do not need to correspond to worldlines of real particles. If
this crucial aspect is not taken into account we may incur in serious
misunderstandings. We observe moreover that the concept of reference frame
fields has been also used since a long time ago by Matolsci [13], although
this author uses a somewhat different terminology.

\section{Physically Equivalent Reference Frames}

The objective of this section is two recall the definition of physically
equivalent reference frames in a spacetime theory and in particular in
\emph{GRT} [1] which will be used in section 6 to prove that the \emph{PLLI}
is false. In order to do that we need to recall some definitions. Let $\langle
M,D,g\rangle$ be a Lorentzian spacetime and let $G_{M}$ be the group of all
diffeomorfisms of $M$, called the \emph{manifold mapping group}. Let
$A\subseteq M$ .

\textbf{Definition 11}. The diffeomorfism $G_{M}$ $\ni$ $h:A\rightarrow M$
induces a deforming mapping
\begin{equation}
h_{\ast}:\mathbf{T}\mapsto h_{\ast}\mathbf{T}=\mathbf{\bar{T}} \label{4.1BIS}%
\end{equation}
such that,

(i) If $f:M\supseteq A\rightarrow\mathcal{R}$, then
\begin{equation}
h_{*}f=f\circ h^{-1}:h(A)\rightarrow\mathcal{R}. \label{4.2BIS}%
\end{equation}

(ii) If $\mathbf{T}\sec T^{(r,s)}(A)\subseteq\sec\mathcal{T}(M)$, where
$T^{(r,s)}(A)$ is the sub-bundle of tensors of the type ($r,s)$ of the tensor
bundle $\mathcal{T}(M)$,then
\begin{align}
&  (h_{*}\mathbf{T})_{he}(h_{*}\omega_{1},...,h_{*}\omega_{r},h_{*}%
X_{1},...,h_{*}X_{s})\nonumber\\
&  \mathbf{T}_{e}(\omega_{1},...,\omega_{r},X_{1},...,X_{s}) \label{4.3BIS}%
\end{align}

$\forall X_{i}\in\sec T_{e}(A),i=1,2,...,r,$ $\forall\omega_{j}\in\sec
T^{*}A,$ $j=1,2,...,s,$ $\forall e\in M$.

(iii) If $D$ is the Levi-Civita connection of $g$ on $M$ and $X,Y\in\sec TM$,
then
\begin{align}
(h_{*}D_{h_{*}X}h_{*}Y)_{he}h_{*}f  &  =(D_{X}Y)_{e}f,\forall e\in
M\nonumber\\
h_{*}D_{h_{*}X}h_{*}Y  &  \equiv h_{*}(D_{X}Y). \label{4.4BIS}%
\end{align}

If $\{f_{\mu}=\partial/\partial x^{\mu}\}$ is a coordinate basis for $TA$ and
$\{\theta^{\mu}=dx^{\mu}\}$ is the corresponding dual basis for $T^{*}A$ and
if
\begin{equation}
\mathbf{T}=T_{\nu_{1}....\nu_{s}}^{\mu_{1}...\mu_{r}}\theta^{\nu_{1}}%
\otimes...\otimes\theta^{\nu_{s}}\otimes f_{\mu_{1}}\otimes...\otimes
f_{\mu_{r}}, \label{4.5BIS}%
\end{equation}

then
\begin{equation}
h_{*}\mathbf{T}=(T_{\nu_{1}....\nu_{s}}^{\mu_{1}...\mu_{r}}\circ h^{-1}%
)h_{*}\theta^{\nu_{1}}\otimes...\otimes h_{*}\theta^{\nu_{s}}\otimes
h_{*}f_{\mu_{1}}\otimes...\otimes h_{*}f_{\mu_{r}}. \label{4.6BIS}%
\end{equation}

Suppose now that $A$ and $h(A)$ can be covered by the local chart
$(U,\varphi)$ of the maximal atlas of $M$, and that $A\subseteq U,$
$h(A)\subseteq U$. Let $\langle x^{\mu}\rangle$ be coordinate functions
associated with $(U,\varphi)$. The mapping
\begin{equation}
x^{\prime\mu}=x^{\mu}\circ h^{-1}:h(U)\rightarrow\mathcal{R} \label{4.7BIS}%
\end{equation}
defines a coordinate transformation $\langle x^{\mu}\rangle$ $\mapsto\langle
x^{\prime\mu}\rangle$ if $h(U)\supseteq A\cup h(A)$. Indeed, $\langle
x^{\prime\mu}\rangle$ are the coordinate functions associated with a local
chart $(V,\chi)$ where $h(U)\subseteq V$ and $U\cap V\neq\emptyset$. Now,
since under these conditions $h_{*}\partial/\partial x^{\mu}=\partial/\partial
x^{\prime\mu}$ and $h_{*}dx^{\mu}=dx^{^{\prime}\mu}$, eqs.(\ref{4.6BIS}) and
(\ref{4.7BIS}) imply that
\begin{equation}
(h_{*}\mathbf{T})_{\langle x^{\prime\mu}\rangle}(he)=\mathbf{T}_{\langle
x^{\mu}\rangle}(e). \label{4.8BIS}%
\end{equation}
In eq.(\ref{4.8BIS}) $\mathbf{T}_{\langle x^{\mu}\rangle}(e)\equiv T_{\nu
_{1}....\nu_{s}}^{\mu_{1}...\mu_{r}}(x^{\mu}(e))$ are the components of
$\mathbf{T}$ in the local coordinate basis $\{\partial/\partial x^{\mu
}\},\{dx^{\mu}\}$ at event $e\in M$, and $(h_{*}\mathbf{T})_{\langle
x^{\prime\mu}\rangle}(he)\equiv\bar{T}_{\nu_{1}....\nu_{s}}^{\prime\mu
_{1}...\mu_{r}}(x^{\prime\mu}(he))$ are the components of $\mathbf{\bar{T}%
}=h_{*}\mathbf{T}$ in the local coordinate basis $\{h_{*}\partial/\partial
x^{\mu}=\partial/\partial x\},\{h_{*}dx^{\mu}=dx^{^{\prime}\mu}\}$ at the
point $he$. Then eq.(\ref{4.8BIS}) reads
\begin{equation}
\bar{T}_{\nu_{1}....\nu_{s}}^{\prime\mu_{1}...\mu_{r}}(x^{\prime\mu
}(he))=T_{\nu_{1}....\nu_{s}}^{\mu_{1}...\mu_{r}}(x^{\mu}(e)). \label{4.9BIS}%
\end{equation}

Using eq.(\ref{4.7BIS}) we can also write
\begin{equation}
\bar{T}_{\nu_{1}....\nu_{s}}^{\prime\mu_{1}...\mu_{r}}(x^{\prime\mu
}(e))=(\Lambda^{-1})_{\alpha_{1}}^{\mu_{1}}...(\Lambda)_{\nu_{s}}^{\beta_{s}%
}T_{\beta_{1}....\beta_{s}}^{\prime\alpha_{1}...\alpha_{r}}(x^{\prime\mu
}(h^{-1}e)) \label{4.10BIS}%
\end{equation}
where $\Lambda_{\alpha}^{\mu}=\partial x^{\prime\mu}/\partial x^{\alpha}$, etc.

\textbf{Definition 12}. Let $h\in G_{M}$. If for a geometrical object
\textbf{T} we have%

\begin{equation}
h_{*}\mathbf{T}=\mathbf{T} \label{4.11BIS}%
\end{equation}
then $h$ is said to be a symmetry of \textbf{T} and the set of all $\{h\in
G_{M}\}$ such that eq.(\ref{4.11BIS}) holds is said to be the symmetry group
of \textbf{T.}

\textbf{Definition 13}. Let $\Upsilon,\Upsilon^{\prime}\in$ Mod $\tau$,
$\Upsilon=(M,D,g,\mathbf{T}_{1},\ldots,\mathbf{T}_{m})$ and $\Upsilon^{\prime
}=(M,D^{\prime},g^{\prime},\mathbf{T}_{1}^{\prime},\ldots,\mathbf{T}%
_{m}^{\prime})$ with the $\mathbf{T}_{i}$, $i=1,\ldots,m$ defined in
$U\subseteq M$ and $\mathbf{T}_{i}^{\prime}$, $i=1,\ldots,m$ defined in
$V\subseteq M$. We say that $\Upsilon$ is equivalent to$\Upsilon^{\prime}$
(and denotes $\Upsilon\sim\Upsilon^{\prime})$ if there exists $h\in G_{M} $
such that $\Upsilon^{\prime}=h_{*}\Upsilon,$ i.e., $V\subseteq h(U)$ and
\begin{equation}
D^{\prime}=h_{*}D,\text{ }g^{\prime}=h_{*}g,\mathbf{T}_{1}^{\prime}%
=h_{*}\mathbf{T}_{1},...,\mathbf{T}_{m}^{\prime}=h_{*}\mathbf{T}_{m}
\label{4.13BIS}%
\end{equation}

Theories satisfying definition 14 are called generally covariant and
$\Upsilon,\Upsilon^{\prime}\in$ Mod $\tau$ represent indeed the same physical model.

\textbf{Definition 14}. Let $\Upsilon,\bar{\Upsilon}\in$ Mod $\tau,$
$\Upsilon=(M,D,g,\mathbf{T}_{1},\ldots,\mathbf{T}_{m})$, $\bar{\Upsilon
}=(M,h_{*}D,h_{*}g,h_{*}\mathbf{T}_{1},\ldots,h_{*}\mathbf{T}_{m})$ with the
$\mathbf{T}_{i}$, $i=1,\ldots,m$ defined in $U\subseteq M$ and $\mathbf{T}%
_{i}^{\prime} $, $i=1,\ldots,m$ defined in $V\subseteq h(U)\subseteq M$ and
such that
\begin{equation}
D=h_{*}D,g=h_{*}g. \label{4.14BIS}%
\end{equation}

Then $\bar{\Upsilon}$ is said to be the $h$-\emph{deformed version }of
$\Upsilon$.

\textbf{Definition 15}. Let $\mathbf{Q}\in\sec TU\subseteq\sec TM,\mathbf{\bar
{Q}}\in\sec TV\subseteq\sec TM$, $U\cap V\neq\emptyset$ and let $\langle
x^{\mu}\rangle$ $,\langle\bar{x}^{\mu}\rangle$ (the coordinate functions
associated respectively to the charts $(U,\varphi)$ and $(V,\bar{\varphi})$)
be respectively a ($nacs|\mathbf{Q}$) and a ($nacs|\mathbf{\bar{Q}}$) and
suppose that $\bar{x}^{\mu}=x^{\mu}\circ\bar{h}^{-1}:\bar{h}(U)\rightarrow
\mathcal{R}$. Thus, $\mathbf{\bar{Q}}=\bar{h}_{*}\mathbf{Q}$ and
$\mathbf{\bar{Q}}$ is said to be a $\bar{h}$-\emph{deformed version} of
$\mathbf{Q}$.

Let $\Upsilon,\bar{\Upsilon}\in$ Mod $\tau$ be as in definition 14. Call
$o=(D,g,\mathbf{T}_{1},\ldots,\mathbf{T}_{m})$ and $\bar{o}=(D,g,h_{*}%
\mathbf{T}_{1},\ldots,h_{*}\mathbf{T}_{m})$. Now, $o$ is such that it solves a
set of differential equations in $\varphi(U)\subset\mathcal{R}^{4}$ with a
given set of boundary conditions denoted $b^{o\langle x^{\mu}\rangle}$, which
we write as%

\begin{equation}
D_{\langle x^{\mu}\rangle}^{\alpha}(o_{\langle x^{\mu}\rangle})_{e}%
=0\ ;\ b^{o\langle x^{\mu}\rangle}\ ;\ e\in U, \label{4.15BIS}%
\end{equation}
and $\bar{o}$ defined in $\bar{h}(U)\subseteq V$ solves
\begin{equation}
D_{\langle\bar{x}^{\mu}\rangle}^{\alpha}(\bar{h}_{*}o_{\langle\bar{x}^{\mu
}\rangle})_{|he}=0\ ;\ b^{\bar{h}_{*}o\langle\bar{x}^{\mu}\rangle}\ ;\ \bar
{h}\text{ }e\in\bar{h}(U)\subseteq V. \label{4.16BIS}%
\end{equation}
In eqs.(\ref{4.15BIS}) and (\ref{4.16BIS}) $D_{\langle x^{\mu}\rangle}%
^{\alpha}$ and $D_{\langle x^{^{\prime}\mu}\rangle}^{\alpha}$ mean
$\alpha=1,2,\ldots,m$ sets of differential equations in $\mathcal{R}^{4}$.

How can an observers living on $M$ discover that $\Upsilon,\bar{\Upsilon}\in$
Mod $\tau$ are deformed versions of each other? In order to answer this
question we need additional definitions.

\textbf{Definition 16}. Let $\mathbf{Q,\bar{Q}}$ be as in definition 15. We
say that $\mathbf{Q}$\textbf{\ }and $\mathbf{\bar{Q}}$ are physically
equivalent according to theory $\tau$ (and we denote $\mathbf{\bar{Q}}%
\sim\mathbf{Q})$ iff%

\begin{equation}
\text{(i)\hspace{0.2in}}D\mathbf{Q}=D\mathbf{\bar{Q}} \label{4.17BIS}%
\end{equation}

and

(ii) the system of differential equations (\ref{4.15BIS}) must have the same
functional form as the system of differential equations (\ref{4.16BIS}) and
$b^{\bar{h}_{*}o\langle\bar{x}^{\mu}\rangle}$ must be relative to $\langle
\bar{x}^{\mu}\rangle$ the same as $b^{o\langle x^{\mu}\rangle}$ is relative to
$\langle x^{\mu}\rangle$ and if $b^{o\langle x^{\mu}\rangle}$ is physically
realizable then $b^{\bar{h}_{*}o\langle\bar{x}^{\mu}\rangle} $ must also be
physically realizable.

\textbf{Definition 17}. Given a reference frame $\mathbf{Q}\in\sec
TU\subseteq\sec TM$ the set of all diffeomorfisms \{$h\in G_{M}\}$ such that
$h_{*}\mathbf{Q}\sim\mathbf{Q}$ forms a subgroup of $G_{M}$ called the
equivalence group of the class of reference frames of kind $\mathbf{Q}$
according to the theory $\tau$.

\textbf{Observation 9}. We can easily verify using definitions 16 and 17 any
two \emph{IRF }in Minkowski space time ($M,D^{\mathbf{\eta}},\mathbf{\eta}$)
(observation 3) are equivalent and that the equivalence group of the class of
inertial reference frames is the Poincar\'{e} group. Of course, we can verify
that the symmetry group (definition 12) of $D^{\mathbf{\eta}}$and
$\mathbf{\eta}$ is also the Poincar\'{e} group. It is the existence of this
symmetry group that permits a mathematical definition of the Special Principle
of Relativity.\footnote{See [14] where we point out that the definition of
physically equivalent reference frames given above leads to contradictions in
\emph{SRT} if superluminal phenomena exist and we insist in mantaining the
validity of the Special Principle of Relativity.} We can also show without
difficulties that two distinct rotating references frames (with have the same
angular velocity relative to a given \emph{IRF} and that have the same radius)
are physically equivalent. Of course, no \emph{IRF} is equivalent to any
rotating frame. A comprehensive example of phenomena related as $\Upsilon
,\bar{\Upsilon}\in$ Mod $\tau$ in definitions 14 and 15 is (in Minkowski
spacetime) the electromagnetic field of a charge at rest relative to an
\emph{IRF} \textbf{I} and the field of a second charge in uniform motion
relative to the same \emph{IRF} $\mathbf{I}$ and its field relative to an
\emph{IRF} $\mathbf{I}^{\prime}$ where the second charge is at rest.

\section{\emph{LLRF}$\gamma$\emph{s} and the Equivalence Principle}

There are many presentations of the \emph{EP} and even very strong criticisms
against it, the most famous being probably the one offered by Synge [15]. We
are not going to bet on this particular issue. Our intention here is to prove
that there are models of \emph{GRT} where the so called Principle of Local
Lorentz Invariance ($\emph{PLLI}$) which according to several authors (see
below) follows from the Equivalence Principle is not valid in general. Our
strategy to prove this strong statement is to give a precise mathematical
wording to the \emph{PLLI} (which formalizes the \emph{PLLI} as introduced by
several authors) in terms of a physical equivalence of \emph{LLRF}$\gamma
$\emph{s} (see below) and then prove that \emph{PLLI }is a false statement
according to \emph{GRT}. We start by recalling formulations and comments
concerning the \emph{EP} and the \emph{PLLI}.

According to Friedmann [16] the ``Standard formulation of the \emph{EP}
characteristically obscure [the] crucial distinction between first order laws
and second order laws by blurring the distinction between infinitesimal laws,
holding at a single point, and local laws, holding on a neighborhood of a point''....

According to our point of view, in order to give a mathematically precise
formulation of Einstein's \emph{EP} besides the distinctions mentioned above
between infinitesimal and local laws, it is also necessary to distinguish
between some very different (but related) concepts, namely, \footnote{These
concepts are in general used without distinction by different authors leading
to misunderstandings and misconceptions.}

(i) The concept of an observer (definition 1);

(ii) The general concept of a reference frame in \emph{GRT} (Definition 4);

(iii) The concept of a natural adapted coordinate system to a reference frame
(Definition 7);

(iv) The concept of \emph{PIRFs} (definition 6) and \emph{LLRF}$\gamma
$\emph{s} (definition 9) on $U\subset M$ ;

(v) The concept of an inertial moving $\emph{observer}$ carrying a tetrad
along $\gamma$ (a geodetic curve), a concept we abbreviate by calling it an
\emph{IMF}$\gamma$ (definition 10).

Einstein's \emph{EP} is formulated by Misner, Thorne and Wheeler (\emph{MTW})
[17] as follows: ``in any and every Local Lorentz Frame (\textbf{LLF}),
anywhere and anytime in the universe, all the (non-gravitational) laws of
physics must take on their familiar special relativistic forms. Equivalently,
there is no way, by experiments confined to small regions of spacetime to
distinguish one \textbf{LLF} in one region of spacetime from any other
\textbf{LLF} in the same or any other region''. \ We comment here that these
authors\footnote{For the best of our knowledge no author gave until now the
fomal definition of a \emph{LLRF} as in definition 9.} did not give a formal
definition of a \textbf{LLF}. They try to make intelligible the \emph{EP} by
formulating its wording in terms of a \emph{LLCC }(see definition 8) and
indeed these authors as many others do not distinguish the concept of a
reference frame $\mathbf{Z}\in\sec TM$ from that of a ($nacs|\mathbf{Z}$).
This may generate misunderstandings. The mathematical formalization of a
\textbf{LLF} \ used by \emph{MTW} (and many other authors) corresponds to the
concept of \emph{LLRF} introduced in definition 9.

In [18] Ciufolini and Wheeler call the above statement of \emph{MTW} the
medium strong form of the \emph{EP}. They introduced also what they called the
strong \emph{EP} as follows: ``in a sufficiently small neighborhood of any
spacetime event, in a locally falling frame, no gravitational effects are
observable''. Again, no mathematical formalization of a locally falling frame
is given, the formulation uses only \emph{LLCC}\footnote{Again, no
mathematical formalization of a locally falling frame is given, the
formulation uses only the concept of \emph{LLCCs}. Worse, if local means in a
neighborhood of a given spacetime event this principle must be false. For,
e.g., it is well known that the Riemann tensor couples locally with spinning
particles. Moreover, the neigbourhood must be at least large enough to contain
an experimental physicist and the devices of his laboratory and must allow for
enough time for the experiments. With a gradiometer builded by Hughes
corporation \ which has an area of\ approximately 400 cm$^{2}$ any one can
easily discover if he is leaving in a region of spacetime with a gravitational
field or if he is living in an accelerated frame in a region of spacetime free
with a zero gravitational field.}.

Following [17,18] recently several authors as, e.g., Will [19], Bertotti and
Grishchuk [20] and Gabriel and Haugan [21] (see also Weinberg [22] claim that
Einstein \emph{EP} requires a sort of\emph{\ }local Lorentz invariance. This
concept is stated in, e.g., [20] with the following arguments.

To start we are told that to state the Einstein \emph{EP} we need to consider
a laboratory that falls freely through an external gravitational field, such
that the laboratory is shielded, from external non-gravitational fields and is
small enough such that effects due to the inhomogeneity of the field are
negligible through its volume. Then, they say, that the local
non-gravitational test experiments are experiments performed within such a
laboratory and in which self-gravitational interactions play no significant
part. They define: ``The Einstein \emph{EP }states that the outcomes of such
experiments are independent of the velocity of the apparatus with which they
are performed and when in the universe they are performed''. This statement is
then called the \emph{Principle of Local Lorentz Invariance} (\emph{PLLI}) and
`convincing' proofs of its validity are offered, and not need to be repeated
here. Prugovecki [23] (pg 62) endorses the \emph{PLLI }and also said that it
can be experimentally verified. In his formulation he translates the
statements of [16-22] in terms of Lorentz and Poincar\'{e} covariance of
measurements done in two different \emph{IMF}$\gamma$\emph{\ }(see Definition
10). Based on these past tentatives of formalization\footnote{See [24] for a
history of the subject.} we give the following one. \medskip

\textbf{Einstein }\emph{EP}: Let $\gamma$ be a timelike geodetic line on the
world manifold $M.$ For any \emph{LLRF}$\gamma$ (see definition 9) all
nongravitational laws of physics, expressed through the coordinate functions
$\langle\xi^{\mu}\rangle$ which are \emph{LLC}\ associated with the
\emph{LLRF}$\gamma$ (definition 9) should at each point along $\gamma$ be
\emph{equal} (up to terms in first order in those coordinates) to their
special relativistic counterparts when the mathematical objects appearing in
these special relativistic laws are expressed through a set of \emph{Lorentz
coordinate functions} naturally adapted to an arbitrary inertial frame
$\mathbf{I\in}\sec TM^{\prime}$, ($M^{\prime}=\mathcal{R}^{4},\mathbf{\eta
},D^{\eta})$ being a Minkowski spacetime (observation 3).

Also, if the \emph{PLLI }would be a true law of nature it could be formulated
as follows:\bigskip

\textbf{Principle of Local Lorentz invariance} (\emph{PLLI}): Any two
\emph{LLRF}$\gamma$ and \emph{LLRF}$\gamma^{\prime}$\ associated with the
timelike geodetic lines $\gamma$ and $\gamma^{\prime}$ of two observers such
that $\gamma\cap\gamma^{\prime}=p$ are physically equivalent at $p$.\medskip

Of course, if \emph{PLLI} is correct, it must follow that from experiments
done by observers inside some \emph{LLRF}$\gamma^{\prime}$--- say
$\mathbf{L}^{\prime}$ that is moving relative to another \emph{LLRF}
$\mathbf{L}$--- there is no means for that observers to determine that
$\mathbf{L}^{\prime}$ is in motion relative to $\mathbf{L}$.

Unfortunately\textbf{\ }the\textbf{\ }\emph{PLLI} is not true. To show that it
is only necessary to find a model of \emph{GRT} where the statement of the
\emph{PLLI} is false. Before proving this result we shall need to prove that
there are models for \emph{GRT} were \emph{PIRFs} are not physically
equivalent also.

\section{Physical non equivalence of \emph{PIRFs }$\mathbf{V}$ and
$\mathbf{Z}$ on a Friedmann Universe}

Recall that \emph{GRT} $\tau_{E}$ is a theory of the gravitational field [4,5]
where a typical model $\tau\in Mod\tau_{E}$ is of the form
\begin{equation}
\tau=<M,g,D,\mathbf{T},(m,\sigma)>, \label{36}%
\end{equation}
where $ST=<M,\mathbf{g},D>$ is a relativistic spacetime and $\mathbf{T}\in
secT^{*}M\otimes T^{*}M$ is called the energy-momentum tensor. $\mathbf{T}$
represents the material and energetic content of spacetime, including
contributions from all physical fields (with exception of the gravitational
field and particles). For what follows we do not need to know the explicit
form of $\mathbf{T}$. The proper axioms of $\tau_{E}$ are:
\begin{equation}
D(\mathbf{g})=0;\quad\mathbf{G}=\mathbf{Ric}-\frac{1}{2}S\mathbf{g}%
=\mathbf{T}, \label{37}%
\end{equation}
where $\mathbf{G}$ is the Einstein tensor, $\mathbf{Ric}$ is the Ricci tensor
and $S$ is the Ricci scalar. The equation of motion of a particle $(m,\sigma)$
that moves only under the influence of gravitation is:
\begin{equation}
D_{\sigma*}\sigma_{*}=0. \label{38}%
\end{equation}
$ST$ is in general not flat, which implies that there do not exist \emph{any}
$\emph{IRF}$ $\mathbf{I}$, i.e., a reference frame such that $D\mathbf{I}=0$.

Now, the physical universe we live in is reasonably represented by metrics of
the Robertson-Walker-Friedmann type [17]. In particular, a very simple
spacetime structure $ST=<M,\mathbf{g},D>$ that represents the main properties
observed (after the big-bang) is formulated as follows: Let $M=\mathcal{R}%
^{3}\times I,I\subset\mathcal{R}$ and $R:I\rightarrow(0,\infty),t\rightarrow
R(t)$ and define $\mathbf{g}$ in $M$ (considering $M$ as subset of
$\mathcal{R}^{4} $) by:
\begin{equation}
\mathbf{g}=dt\otimes dt-R(t)^{2}\sum dx^{i}\otimes dx^{i},i=1,2,3. \label{39}%
\end{equation}

Then $\mathbf{g}$ is a Lorentzian metric in $M$ and $\mathbf{V}=\partial
/\partial t$ is a time-like vector field in $(M,G)$. Let $<M,g,D>$ be oriented
in time by $\partial/\partial t$ and spacetime oriented by $dt\wedge
dx^{1}\wedge dx^{2}\wedge dx^{3}$. Then $<M,g,D>$ is a relativistic spacetime
for $I=(0,\infty)$.

Now, $\mathbf{V}=\partial/\partial t$ is a reference frame. Taking into
account that the connection coefficients in a ($nacs$%
$\vert$%
$\mathbf{V}$) given by the coordinate system in eq.(\ref{39}) are
\begin{align}
\Gamma_{kl}^{i}  &  =0,\text{ }\Gamma_{kl}^{0}=R\dot{R}\delta_{kl}\text{,
}\Gamma_{0l}^{k}=\frac{\dot{R}}{R}\delta_{l}^{k}\nonumber\\
\Gamma_{00}^{i}  &  =\Gamma_{0l}^{0}=\Gamma_{00}^{0}=0, \label{44}%
\end{align}
we can easily verify that \textbf{V} is a \emph{PIRF (}according to definition
6) since $D_{\mathbf{V}}\mathbf{V}=0$ and $\ d\alpha_{\mathbf{V}}\wedge
\alpha_{\mathbf{V}}=0$, \ $\alpha_{\mathbf{V}}=\mathbf{g}(\mathbf{V},)$. Also,
since $\alpha_{\mathbf{V}}=dt$, \textbf{V} is proper time synchronizable.

\textbf{Proposition 2}\footnote{The suggestion of the validity of a
proposition like the one formalized by proposition 3 has been first proposed
by Rosen [25]. However, he has not been able to identify the true nature of
the $\mathbf{V}$ and $\ \mathbf{Z}$ which he thought as representing
`inertial' frames. He tried to show the validity of the proposition by
analyzing the output of mechanical and optical experiments done inside the
frames \textbf{V} and \textbf{Z. } We present in section 7.3 a simplified
version of his suggested mechanical experiement. It is important to emphasize
here that from the validity of the proposition 3 he suggested that it implies
in a breakdown of the \emph{PLLI}. Of course, the \emph{PLLI} refers to the
physical equivalence of \emph{LLRF}$\gamma s$. Also the proof of proposition 3
given above is original.}. In a spacetime defined by Eq.(\ref{4.16BIS}) which
is a model of $\tau_{E}$ there exists a \emph{PIRF} $\mathbf{Z}\in\sec TU$
which is not physically equivalent to $\mathbf{V}=\partial/\partial t$.

\textbf{Proof}: Let $\mathbf{Z}\in\sec TU$ be given by
\begin{equation}
\mathbf{Z}=\frac{(R^{2}+u^{2})^{1/2}}{R}\partial/\partial t+\frac{u}{R^{2}%
}\partial/\partial x^{1} \label{45}%
\end{equation}
where in eq.(\ref{45}) $u\neq0$ is a real constant.

Since $D_{\mathbf{Z}}\mathbf{Z}=0$ and $\ d\alpha_{\mathbf{Z}}\wedge
\alpha_{\mathbf{Z}}=0$, \ $\alpha_{\mathbf{Z}}=\mathbf{g}(\mathbf{Z},)$, it
follows that $\mathbf{Z}$\textbf{\ }is a \emph{PIRF}\footnote{Introducing the
(nacs%
$\vert$%
\textbf{Z}) given by eq.(\ref{48}) we can show that $\alpha_{\mathbf{Z}%
}=dt^{\prime}$ and it follows that is also proper time synchronizable.}. All
that is necessary in order to prove our proposition is to show that
$D\mathbf{Z}\neq D\mathbf{V}$. It is enough to prove that the expansion ratios
$\mathbf{\Theta}_{\mathbf{Z}}\neq\mathbf{\Theta}_{\mathbf{Z}}$. Indeed,
eq.(\ref{2}) gives
\begin{align}
\mathbf{\Theta}_{\mathbf{V}}  &  =3\dot{R}/R,\nonumber\\
\mathbf{\Theta}_{\mathbf{Z}}  &  =\frac{\left[  R\dot{R}+2\dot{R}(R^{2}%
+u^{2})^{1/2}\right]  }{R^{2}\left(  R^{2}+u^{2}\right)  ^{1/2}}, \label{45.n}%
\end{align}
where
\begin{equation}
v=\left.  R(\frac{d}{dt}x^{1}\circ\gamma)\right\vert _{t=0}=u(1+u^{2})^{-1/2}
\label{45bis}%
\end{equation}
is the initial metric velocity of $\mathbf{Z}$ relative to $\mathbf{V}$, since
we choose in what follows the coordinate function $t$ such that $R(0)=1$,
$t=0$ being taken as the present epoch where the experiments are done. Then,
$\mathbf{\Theta}_{\mathbf{V}}(p_{0})=3a,$ and for $v<<1,$ $\mathbf{\Theta
}_{\mathbf{Z}}(p_{0})=3a-av^{2}]$.$\blacksquare$

\subsection{Mechanical experiments distinguish \emph{PIRFs}}

\textbf{\ }If accepted, the \emph{PLLI} says that \emph{LLRF}$\gamma$\emph{s
}at $p\in M$ are physically equivalent and that there are no mechanical
experiments that can distinguish between them. We shall prove below that
\emph{PLLI} is false, at least, if one of these experiments refers to the
measurement of the expansion ratio of the \emph{LLRF}$\gamma$\emph{s }at $p\in
M$.

The question arises: can mechanical experiments (distinct from the one
designed to measure the expansion ratio) distinguish between the \emph{PIRFs}
\textbf{V }and \textbf{Z? }The answer is yes. To prove our statement we
proceed as follows.

(i) We start by finding a $(nacs|\mathbf{Z})$. To do that we note if $\gamma$
is an integral curve of $\mathbf{Z}$, we can write
\begin{equation}
\mathbf{Z}_{|\gamma}=[\frac{d}{ds}(x^{\mu}\circ\gamma)\frac{\partial}{\partial
x^{\mu}}]_{|\gamma} \label{46}%
\end{equation}
where \ $s$ is the proper \ time parameter along $\gamma$. Then, we can write
[taking into account eqs.(\ref{44})] its parametric equations as
\begin{equation}
\frac{d}{dt}x^{1}\circ\gamma=\frac{(\frac{d}{ds}x^{1}\circ\gamma)}{(\frac
{d}{ds}t^{\ }\circ\gamma)}=\frac{u}{R(R^{2}+u^{2})^{1/2}};\quad x^{2}%
\circ\gamma=0;\quad x^{3}\circ\gamma=0 \label{47}%
\end{equation}
(The direction $x^{1}\circ\gamma=0$ is obviously arbitrary). We then choose
for $(nacs|\mathbf{Z})$ the coordinate functions $(t^{\prime},x^{1^{\prime}%
},x^{2^{\prime}},x^{3^{\prime}})$ given by:
\begin{equation}
x^{1^{\prime}}=x^{1}-u\int_{0}^{t}dr\frac{1}{R(r)[R(r)^{2}+u^{2}]^{1/2}};\quad
x^{2^{\prime}}=x^{2};\nonumber
\end{equation}
\begin{equation}
x^{3^{\prime}}=x^{3};\quad t^{\prime}=\int_{0}^{t}dr\frac{[R(r)^{2}%
+u^{2}]^{1/2}}{R(r)}-ux^{1} \label{48}%
\end{equation}
We then get:
\begin{equation}
\mathbf{g}=dt^{\prime}\otimes dt^{\prime}-\overline{R}(t^{\prime})^{2}\left\{
\begin{array}
[c]{c}%
\left[  \frac{1-v^{2}(1-\overline{R}(t^{\prime})^{-2})}{1-v^{2}}\right]
dx^{1^{\prime}}\otimes dx^{1^{\prime}}\\
+dx^{2^{\prime}}\otimes dx^{2^{\prime}}+dx^{3^{\prime}}\otimes dx^{3^{\prime}}%
\end{array}
\right\}  , \label{49}%
\end{equation}
and the connection coefficients in the ($nacs|\mathbf{Z}$) are,
\begin{align}
\text{ }\bar{\Gamma}_{kl}^{0}  &  =\frac{\overset{.}{\bar{R}}\bar{R}^{2}%
}{(\bar{R}^{2}+u^{2})^{\frac{1}{2}}}\text{ }\delta_{kl}\text{, }\bar{\Gamma
}_{01}^{1}=\frac{\overset{.}{\bar{R}}\bar{R}^{2}}{(\bar{R}^{2}+u^{2}%
)^{\frac{3}{2}}}\text{, }\bar{\Gamma}\text{ }_{02}^{2}=\bar{\Gamma}\text{
}_{03}^{3}=\frac{\overset{.}{\bar{R}}}{(\bar{R}^{2}+u^{2})^{\frac{1}{2}}%
},\text{ }\nonumber\\
\bar{\Gamma}_{kl}^{i}  &  =0\text{, }\bar{\Gamma}_{00}^{i}=\bar{\Gamma}\text{
}_{0l}^{0}=\bar{\Gamma}\text{ }_{00}^{0}=0. \label{50}%
\end{align}
where $\overline{R}(t^{\prime})=R(t(t^{\prime}))$ and $v$ given by
eq.(\ref{45bis}) is the initial metric velocity of $\mathbf{Z}$ relative to
$\mathbf{V}$, since we choose in what follows the coordinate function $t$ such
that $R(0)=1 $, $t=0$ being taken as the present epoch where the experiments
are done. $\mathbf{Z}=\partial/\partial t^{\prime}$ is a proper time
synchronizable reference frame and we can verify that $t^{\prime}$ is the time
shown by standard clocks at rest in the $\mathbf{Z}$ frame synchronized \`{a}
l'Einstein. Notice that an observer at rest in $\mathbf{Z}$ does \emph{not}
know a priori the value of $v$. \ He can discover this value as
follows:\medskip

(ii) The solution of the equation of motion for a free particle $(m,\sigma)$
in $\mathbf{V}$ with the initial conditions at $p_{0}=(0,$ $x^{i}\circ
\sigma(0)=0),$ $i=1,2,3$ and $\frac{d}{dt}x^{i}\circ\sigma(0)=\bar{u}^{i}$ for
a fixed $i$ and $\frac{d}{dt}x^{i}\circ\sigma(0)=0,j\neq i$, is given by an
equation analogous to Eq.(\ref{47}). The accelerations are such that
\begin{equation}
\left.  \frac{d^{2}}{ds^{2}}x^{j}\circ\sigma(t))\right|  _{p_{0}}=0,\text{
}j\neq i. \label{51}%
\end{equation}
(iii) The equation of motion for a free particle $(m,\sigma^{\prime})$ in
$\mathbf{Z}$ , can be write as (we write for simplicity in what follows
$\frac{d^{2}}{ds^{2}}x^{\prime1}\circ\sigma^{\prime}(t^{\prime})\equiv
\frac{d^{2}}{ds^{2}}x^{\prime1}(t^{\prime})\equiv\frac{d^{2}}{ds^{2}}%
x^{\prime1}$, etc...)
\begin{align}
\frac{d^{2}x^{\prime1}}{ds^{2}}  &  =-2\frac{\overset{.}{\bar{R}}\bar{R}^{2}%
}{(\bar{R}^{2}+u^{2})^{\frac{3}{2}}}\frac{dx}{dt^{\prime}}^{\prime1}%
(\frac{dt^{\prime}}{ds})^{2},\nonumber\\
\frac{d^{2}x^{\prime i}}{ds^{2}}  &  =-2\frac{\overset{.}{\bar{R}}}{(\bar
{R}^{2}+u^{2})^{\frac{1}{2}}}\frac{dx}{dt^{\prime}}^{\prime i}(\frac
{dt^{\prime}}{ds^{2}}),\text{ }i=2,3,\nonumber\\
\frac{d^{2}t}{ds^{2}}^{\prime}  &  =-2\frac{\overset{.}{\bar{R}}\bar{R}^{2}%
}{(\bar{R}^{2}+u^{2})^{\frac{1}{2}}}\text{ }\left[  \left(  \frac{dx^{\prime
1}}{dt^{\prime}}\right)  ^{2}+\left(  \frac{dx^{\prime2}}{dt^{\prime}}\right)
^{2}+\left(  \frac{dx^{\prime3}}{dt^{\prime}}\right)  ^{2}\right] \nonumber\\
\frac{dt^{\prime}}{ds}  &  =\left[  1+\widehat{\overline{R}}^{2}\left(
\frac{dx^{\prime1}}{dt^{\prime}}\right)  ^{2}+\bar{R}^{2}\left(
\frac{dx^{\prime2}}{dt^{\prime}}\right)  ^{2}+\bar{R}^{2}\left(
\frac{dx^{\prime3}}{dt^{\prime}}\right)  ^{2}\right]  ^{-\frac{1}{2}}
\label{52}%
\end{align}
where the dot over $R$ in eq.(\ref{52}) means derivative with respect to
$t^{\prime}$ and $\widehat{\overline{R}}$ denotes the square root of the
coefficient of $dx^{1^{\prime}}\otimes dx^{1^{\prime}}$ term in eq.(\ref{49}).

>From these equations it is easy to verify that the two situations :

(a) motion in the $(x^{1^{\prime}},x^{2^{\prime}})$ plane with initial
conditions at $p_{0}$ with coordinates $(t^{\prime}=0,x^{1^{\prime}%
}=x^{2^{\prime}}=0=x^{3^{\prime}})$ given by
\begin{equation}
\left.  \frac{dx^{1^{\prime}}(t^{\prime})}{dt^{\prime}}\right|  _{p_{_{0}}%
}=v_{1}^{\prime},\left.  \frac{dx^{2^{\prime}}(t^{\prime})}{dt}\right|
_{p_{_{0}}}=0,\quad\label{53}%
\end{equation}
and

(b) motion in the $(x^{1^{\prime}},x^{2^{\prime}})$ plane with initial
conditions at $p_{0}$ with coordinates $(t^{\prime}=0,x^{1^{\prime}%
}=x^{2^{\prime}}=0=x^{3^{\prime}})$ given by
\begin{equation}
\left.  \frac{dx^{1^{\prime}}(t^{\prime})}{dt^{\prime}}\right|  _{p_{0}%
}=0,\quad\left.  \frac{dx^{2^{\prime}}(t^{\prime})}{dt^{\prime}}\right|
_{p_{0}}=v_{2^{\prime}}, \label{54}%
\end{equation}
produce asymmetrical outputs for the measured accelerations along $x^{\prime
1}$ and $x^{\prime2}$. The explicit values depends of course of the function
$R(t)$. If we take $R(t)=1+at$, the asymmetrical accelerations will be given
in terms of $a<<1$ and $v$. This would permit in principle for the
\emph{eventual} observers living in the \emph{PIRF} $\mathbf{Z}$ to infer the
value of $u$ (or $v$).$\bigskip$

\section{\emph{LLRF}$\gamma$ and \emph{LLRF}$\gamma^{\prime} $\ are not
Physically Equivalent on a Friedmann Universe.}

\textbf{Proposition 3}. There are models of \emph{GRT} for which two
\emph{Local Lorentz Reference Frames }are not physically equivalent.\medskip

\textbf{Proof}: Take as model of \emph{GRT} the one just described above where
$\mathbf{g}$ is given by eq.(\ref{39}) and take as before, $R(t)=1+at$.
Consider two integral lines $\gamma$ and $\gamma^{\prime}$ of $\mathbf{V}$ and
$\mathbf{Z}$ such that $\gamma$ $\cap$ $\gamma^{\prime}=p$.

We can associate with these two integral lines the \emph{LLRF}$\gamma
$\emph{\ }$\mathbf{L}$ and the \emph{LLFR}$\gamma^{\prime}$ $\mathbf{L}%
^{\prime}$ as in definition 9. Observe that $\left.  \mathbf{V}\right|
_{\gamma}\mathbf{=}\left.  \mathbf{L}\right|  _{\gamma}$ and $\left.
\mathbf{Z}\right|  _{\gamma^{\prime}}\mathbf{=}\left.  \mathbf{L}^{\prime
}\right|  _{\gamma^{\prime}}$.

Definition 17 says that if $\mathbf{L}$\textbf{\ }and $\mathbf{L}^{\prime}$
are physically equivalent then we must have $D\mathbf{L}=D\mathbf{L}^{\prime}%
$. However, a simple calculation shows that in general $D\mathbf{L}\neq
D\mathbf{L}^{\prime}$ even at $p$! Indeed, we have
\begin{equation}
\mathbf{\Theta}_{\mathbf{L}}=-3t\left(  \frac{\dot{R}}{R}\right)  ^{2},
\label{nova1}%
\end{equation}
\begin{align}
\mathbf{\Theta}_{\mathbf{L}^{\prime}}  &  =2\overset{.}{\bar{R}}(\bar{R}%
^{2}+u^{2})^{1/2}+\dot{R}-\frac{\overset{.}{\bar{R}\bar{R}^{2}}}{(\bar{R}%
^{2}+u^{2})^{3/2}}-\frac{2\overset{.}{\bar{R}}}{(\bar{R}^{2}+u^{2})^{1/2}%
}-\frac{2\overset{.}{\text{ }\bar{R}^{2}}\bar{R}^{4}}{(\bar{R}^{2}+u^{2})^{3}%
}tx^{\prime1}\nonumber\\
&  -\frac{2\overset{.}{\text{ }\bar{R}^{2}}}{(\bar{R}^{2}+u^{2})}tx^{\prime
2}\text{ }-\frac{2\overset{.}{\text{ }\bar{R}^{2}}}{(\bar{R}^{2}+u^{2}%
)}tx^{\prime3}. \label{nova2}%
\end{align}

>From equations (\ref{nova1}) and (\ref{nova2}) we see that the expansions
ratios $\Theta_{\mathbf{L}}$ and $\Theta_{\mathbf{L}^{\prime}}$ are different
in our model and then it follows our result. At $p$, we have $\mathbf{\Theta
}_{\mathbf{L}}(p)=0\ $and $\mathbf{\Theta}_{\mathbf{L}^{\prime}}(p)=2av^{2}%
$.$\blacksquare\medskip$

\textbf{Observation 10}. Proposition 3 establishes that in a Friedmann
universe there is a \emph{LLRF}$\gamma$ (say $\mathbf{L}$) whose expansion
ratio at $p$ is zero. Any other \emph{LLRF}$\gamma^{\prime}$ (say
$\mathbf{L}^{\prime}$) at $p$ will have an expansion ratio at $p$ given by
$2av^{2}$, where $a\ll1$ and $v$ is the metric velocity of $\mathbf{L}%
^{\prime}$ relative to $\mathbf{L}$ at $p$. This expansion ratio can in
principle be measured and this is the reason for the nonvalidity of the
\emph{PLLI} as formulated by many contemporary physicists and formalized
above. Note that all experimental verifications of the \emph{PLLI} mentioned
by the authors that endorse the \emph{PLLI} have been obtained for
\emph{LLRF}$\gamma s$ moving with $v<<1$, and have no accuracy in order to
contradict the result we found. We do not know of any experiment that has been
done on a \emph{LLRF}$\gamma$ which enough precision to verify the effect .
Anyway the non physical equivalence between $\mathbf{L}$ and $\mathbf{L}%
^{\prime}$ is a \emph{prediction} of \emph{GRT} and must be accepted if this
theory is right. \medskip\textit{PLLI} is only approximately valid.

We conclude this section by recalling that Friedman [16] formulates the
\emph{PLLI} by saying that if $<U,\xi^{\mu}>,<U^{\prime},\bar{\xi}^{\mu}>$
($U\cap U\neq\emptyset$) are \emph{LLCC } adapted to the \textbf{L }and
$\mathbf{L}^{\prime}$ respectively, then the \emph{PLLI }implies that two
experiments whose initial conditions read alike in terms of $<\xi^{\mu}>$ and
$<\bar{\xi}^{\mu}>$ will also have the same \emph{outcome} in terms of these
coordinate charts.

Friedman's statement is not correct, of course, in view of proposition 3
above, for measurement of the expansion ratio of a reference frame is
something objective and, of course, it is a physical experiment. However, for
experiments different from this one of measuring the expansion ratio we can
accept Friedman's formulation of the \emph{PLLI} as an approximately true statement.

\textbf{Observation 11}.\ Recall the expansion ratios calculated for
$\mathbf{V,Z,L,L}^{\prime}$. Now, $a<<1$. Then, if $v<<1$ the \emph{LLRF}%
$\gamma$ \textbf{L} and the \emph{LLRF}$\gamma^{\prime}$ $\mathbf{L}^{\prime}$
will be almost `rigid ' whereas the $\mathbf{V}$ and\textbf{\ }$\mathbf{Z}$
\ are expanding. In other words, the $\mathbf{L}$ and $\mathbf{L}^{\prime}$
frames can be thought as being physically materialized in their domain by real
solid bodies and thus correspond to small real laboratories, the one used by
physicists. On the other hand it is well known that the $\mathbf{V}$ frame is
an idealization, since only the center of mass of the galactic clusters are
supposed to be comoving with the $\mathbf{V}$ frame, i.e., each center of mass
of a galactic cluster follows some particular integral line of $\mathbf{V}$.
Concerning the $\mathbf{Z}$ frame, in order for it to be realized as a
physical system it must be build with a special matter that suffers in all
points of its domain an expansion a little bit greater than the cosmic
expansion. Of course, such a frame would be a very artificial one, and we
suspect that such a special matter cannot be prepared in our universe.

\section{Conclusions}

In this paper we presented a careful analysis of the concept of a reference
frames in \emph{GRT} which are modelled as certain unit timelike vector fields
and gave a physically motivated and mathematical rigorous definition of
physically equivalent reference frames. We investigate which are the reference
frames in \emph{GRT} which share some of the properties of the inertial
reference frames of \emph{SRT}. We found that in \emph{GRT} there are two
classes of frames that appears as generalizations of the inertial frames of
\emph{SRT}. These are the class of the pseudo inertial reference frames
(\emph{PIRFs}) \ and the class of the Local Lorentz reference frames
(\emph{LLRF}$\gamma$\emph{s)} . We showed that \emph{LLRF}$\gamma$\emph{s} are
\emph{not} physically equivalent in general and this implies that the so
called Principle of Local Lorentz invariance (\emph{PLLI}) which several
authors state as meaning that \emph{LLRF}$\gamma$\emph{s} are equivalent is
false. It can only be used as an approximation in experiments that do not have
enough accuracy to measure the effect we found. We prove moreover that there
are models of \emph{GRT} where \emph{PIRFs }are not physically equivalent also.

\begin{description}
\item \textbf{Acknowledgments:}The authors are grateful to Professors U.
Bartocci, J. Vaz, Jr. and A. Saa and Drs. A. M. Moya and D. S. Thober for
stimulating discussions and to the referees' comments which helped to improve
the quality of the paper\medskip
\end{description}

{\large References}

\hspace{0.13in}

[1] W. A. Rodrigues, Jr., M. Scanavini and L. P. de Alcantara , \ \emph{Found.
Phys. Lett.} \textbf{3}, 59 (1990).

[2] N. Bourbaki, \emph{Theorie des Ensembles}, Hermann, Paris, 1957.

[3] H. Reichenbach, \emph{The Philosophy of Space and Time}, Dover, New York, 1958.

[5] W. A. Rodrigues, Jr., and M. A. F. Rosa, \emph{Found. Phys}. \textbf{19,
}705 (1989).

[4] R. K. Sachs and H. Wu, \emph{General Relativity for Mathematicians},
Springer Verlag, N. York, Berlin, 1977.

[6] W. A. Rodrigues, Jr., and de E.C. Oliveira, \emph{Phys. Lett.A
}\textbf{140}, 479 (1989).

[7] Y. Choquet-Bruhat, C. Dewitt-Morette and M. Dillard-Bleick,
\emph{Analysis, Manifolds and Physics} (revised version), North Holland Pub.
Co., Amsterdam, 1982.

[8] W. A. Rodrigues, Jr. and M. Sharif, \emph{Rotating Frames in SRT}:
\emph{Sagnac Effect and Related Issues}, \emph{Found. Phys. }\textbf{31}, 1767 (2001).

[9] S. Kobayashi and K. Nomizu, \emph{Foundations of Differential Geometry},
vol. \textbf{1}, J. Wiley \& Sons, New York, 1963.

[10] R. L. Bishop and S. I. Goldberg,\emph{\ Tensor Analysis on Manifolds},
Dover Publ. Inc., 1980.

[11] A. S. Eddington, \emph{The Mathematical Theory of Relativity} (3rd
edition), Chelsea Publ. Co., New York, 1975.

[12] R. C. Tolman, \emph{Relativity, Thermodynamics and Cosmology}, Dover
Publ. Inc., New York, 1987.

[13] T. Matolsci, \emph{Spacetime Withouth Reference Frames}, Akad\'{e}miai
Kiad\'{o}, Budapest, 1993.

[14] J. E. Maiorino, and W. A. Rodrigues, Jr., What is Superluminal Wave
Motion?, \emph{Sci. and Tech. Mag}. \textbf{4}(2) (1999), see
{\footnotesize http://www.ime.unicamp.br/ rel\_pesq/1999/rp59-99.html}

[15] J. L. Synge, \emph{Relativity: The General Theory}, North Holland,
Amsterdam, 1960.

[16] M. Friedman, \emph{Foundations of Spacetime Theories}, Princeton Univ.
Press, Princeton (1983).

[17] C. M. Misner, K. S. Thorne and J. A. Wheeler, \emph{Gravitation}, W.H.
Freeman and Co. San Francesco, 1973.

[18] I. Ciufuolini and J. A. Wheeler, \emph{Gravitation and Inertia},
Princeton University Press, Princeton, N. Jersey, 1995.

[19] C. M. Will, \emph{Theory and Experiment in Gravitational Physics},
Cambridge Univ. Press, Cambridge, 1980.

[20] B. Bertotti and L.P. Grishchuk, $\emph{Class.Quantum}$ \emph{Gravity}
\textbf{7}, 1733 (1990).

[21] M. D. Gabriel and M.P. Haugan, \emph{Phys. Rev. D} \textbf{141}, 2943 (1990).

[22] S. Weinberg, \emph{Gravitation and Cosmology}, J. Wiley \& Sons, N. York, 1972.

[23] E. Prugrovecki, \emph{Quantum Geometry: A Framework Quantum General
Relativity}, Kluwer Acad. Pub., Dordrecht, 1992.

[24] J. Norton in D. Howard and J. Stachel (eds.), \emph{Einstein and the
History of General Relativity}, Birkhauser, Boston, 1989.

[25] N. Rosen, \emph{Proc. Israel Acad. Sci. and Hum}. \textbf{1}, 12 (1968).
\end{document}